\documentclass[aps,prb,twocolumn,floatfix,amsmath,amssymb]{revtex4}
\usepackage[final]{graphicx}
\usepackage{bm}
\usepackage{float}
\usepackage{afterpage}
\newcommand\vex[1]{\mathbf{#1}}

\newcommand{\bk}{{\bf k}}

\newcommand{\br}{{\bf r}}

\newcommand{\bA}{{\bf A}}

\begin{document}
\title{Interaction driven instabilities of a Dirac semi-metal}
\author{C. Weeks}
\affiliation{Department of Physics and Astronomy, University of British Columbia, Vancouver, BC, Canada V6T 1Z1}
\author{M. Franz}
\affiliation{Department of Physics and Astronomy, University of British 
Columbia, Vancouver, BC, Canada V6T 1Z1}
\begin{abstract}
We explore the possible particle-hole instabilities that can arise in a system of massless Dirac fermions on both the honeycomb and  $\pi$-flux square lattices with short range interactions. Through analytical and numerical studies we show that these instabilities can result in a number  of interesting phases. In addition to the previously identified charge and spin density wave phases and  the exotic `quantum anomalous Hall' (Haldane)  phase, we establish the existence of the dimerized `Kekul\'e' phase over a significant portion of the phase diagram and discuss the possibility of its spinful counterpart, the `spin Kekul\'e' phase. On the $\pi$-flux square lattice we also find various stripe phases, which do not occur on the honeycomb lattice. The Kekul\'e phase is described by a Z$_3$ order parameter whose singly quantized vortices carry fractional charge $\pm e/2$. On the $\pi$-flux lattice the analogous dimerized phase is described by a Z$_4$ order parameter. We perform a fully self-consistent calculation of the vortex structure inside the dimerized phase and find that close to the core the vortex resembles a familiar superconducting U(1) vortex, but at longer length scales a clear Z$_4$ structure emerges with domain walls along the lattice diagonals.
\end{abstract}

\maketitle
\section{\label{intro}Introduction}
Electrons hopping on the two-dimensional honeycomb lattice and the square lattice with a half magnetic flux quantum piercing each plaquette ($\pi$-flux lattice) exhibit, near half filling, a linearly dispersing excitation spectrum characteristic of massless Dirac fermions. The honeycomb lattice is realized in single layer graphene\cite{graphene} while the $\pi$-flux square lattice has the potential to be realized in artificially engineered semiconductor heterostructures.\cite{berciu1} Such massless Dirac fermions exhibit a host of fascinating properties which underlie much of the current interest in graphene and the related systems. 

Although Dirac fermions in graphene are intrinsically massless, it is interesting to contemplate the effects on its electronic properties  of a bandgap that would give rise to {\em massive} Dirac fermions. Experimentally, such a bandgap can be realized by placing graphene on a specific substrate\cite{lanzara1} that breaks the sublattice symmetry. Even more interesting is the possibility of the interaction-driven gap, resulting in a Mott insulating behavior. Aside from the conventional charge density wave (CDW) and spin density wave (SDW) instabilities, the structure of graphene allows for more interesting phases such as the quantum anomalous Hall (QAH) phase discussed by Haldane\cite{haldane1} and the quantum spin Hall (QSH) phase.\cite{zhang1,kane1} These phases are characterized by nontrivial topological invariants\cite{tknn,kane2,moore1} and exhibit the quantum Hall effect in absence of magnetic field and the quantum spin-Hall effect, respectively. Although, as far as we know, these phases do not occur in natural graphene, the QSH phase was predicted to occur\cite{bernevig1} and subsequently observed\cite{konig1} in HgTe quantum wells. Also, theoretical studies of these phases led to the pioneering work on topological insulators in two and three spatial dimensions.\cite{kane2, moore1, roy1, fu1}

Another interesting gapped phase on the honeycomb lattice is the Kekul\'e phase, illustrated in Fig.\ 1b. The Kekul\'e phase is topologically trivial but it is characterized by a Z$_3$ order parameter. The latter describes three degenerate Kekul\'e ground states, obtained by translating the pattern depicted in Fig.\ 1b by the Bravais lattice primitive vectors. Singly quantized vortices in the phase of this order parameter have been shown to possess a stable fermionic zero mode, carry fractional charge\cite{HouChaMud07a} $\pm e/2$, and obey fractional exchange statistics.\cite{SerFra07x,ryu1}  Similar physics is realized in the dimerized phase on the $\pi$-flux square lattice.\cite{Ser08a} 
\begin{figure}
\includegraphics[width = 8.0cm]{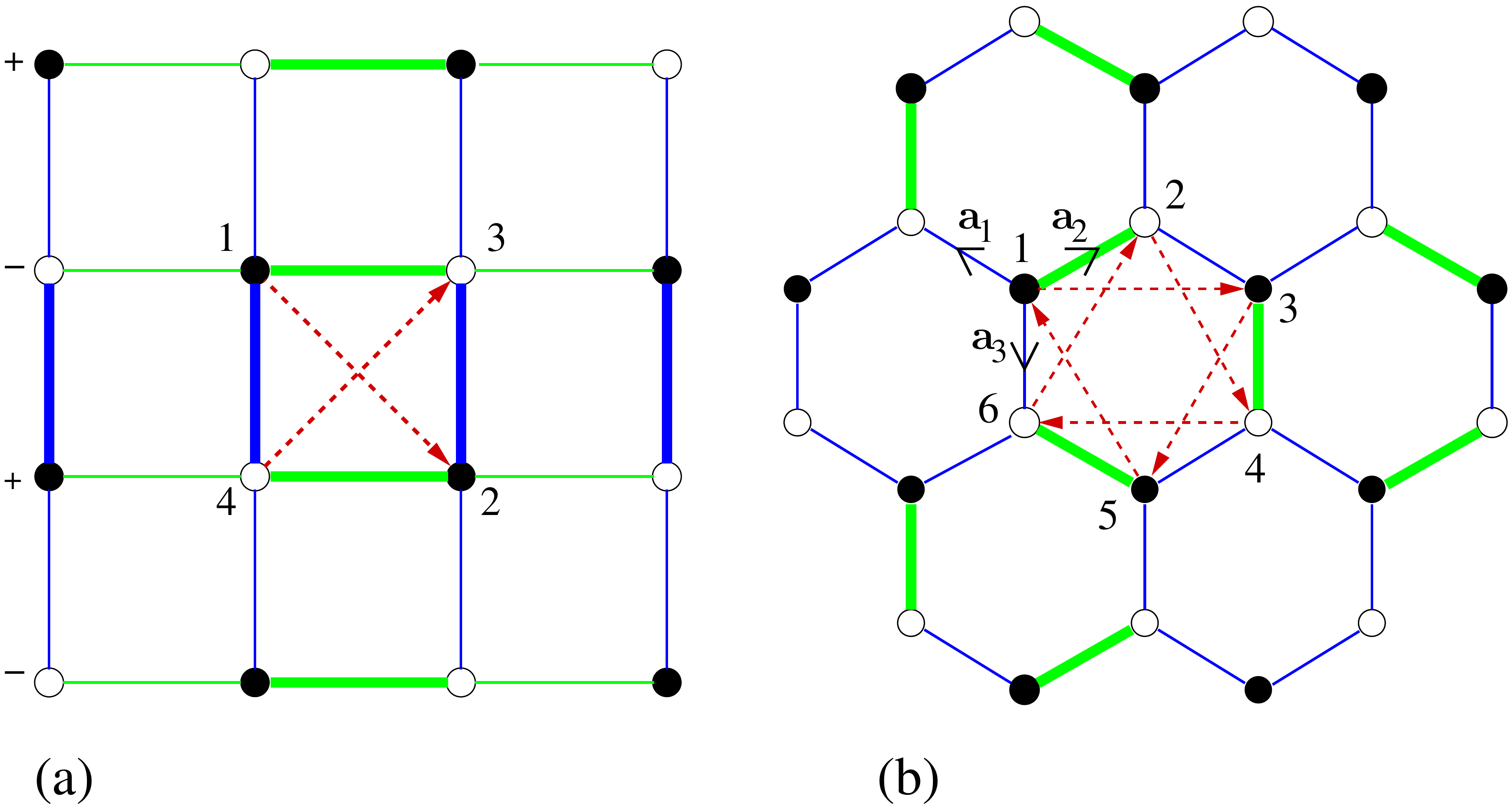}
\caption{The model: (a) Square lattice with $\frac12\Phi_0$ 
magnetic flux per plaquette and dimerized hopping amplitudes. The $\pm$ on the 
left for each row show the choice of gauge for the Peierls phase factors. The thick (thin) bonds indicate an increased (decreased) hopping amplitude in the $\hat x$ (green) and $\hat y$ (blue) directions, the $\bullet$/$\circ$ dots on the lattice sites represent an increase/decrease in local charge density, and the dotted red line indicates the NNN hopping (showing only inside a single plaquette). The four sites of the 
unit cell are marked 1-4. (b) Honeycomb lattice with similar color scheme}
 \label{lattice}
\end{figure}

In order to map out these interesting phases, Raghu {\em et al.}\cite{Raghu} studied a simple model describing
fermions, both spinless and spinful, on the honeycomb lattice with short range interactions. They found phase diagrams containing CDW, SDW, QAH and QSH order but did not consider the possibility of the Kekul\'e phase. With the goal of complementing this previous work we study similar models and find that the Kekul\'e phase is in fact present over a large portion of the phase diagram.  Specifically, for spinless fermions, Kekul\'e order is established when the nearest-neighbor (NN) and next-nearest-neighbor (NNN) repulsion parameters $V_1$ and $V_2$ are of similar size. On the $\pi$-flux square lattice we obtain a phase diagram with the Kekul\'e phase replaced by a simpler dimerized phase having similar properties. In addition, we find stripe phases that tend to dominate over the QAH and the dimerized phases when $V_2$ is large. Nevertheless, when the third-neighbor repulsion $V_3$ is included, the Kekul\'e and QAH phases are stabilized over a portion of the phase diagram. For spinful fermions, the possibility of a novel spin-Kekul\'e phase characterized by the emergence of spatially modulated spin-dependent hopping amplitude is explored.

Having a mean-field theory of the dimerized phase for the $\pi$-flux model 
allows us to perform a self-consistent calculation of the electronic structure for a system having a vortex in the dimer order parameter, which is set up in a fashion similar to Ref.\ \onlinecite{Ser08a}. The underlying $Z_{4}$ symmetry of the lattice raises the issue of whether it is realistic to talk of $U(1)$ vortices in such a system, but as we have shown previously the 4-fold anisotropy induced by the lattice is very weak and should support such vortices on length scales long compared to the lattice spacing. Further evidence is supplied here to support this claim.

\section{\label{model}Spinless fermions}
We begin with the tight-binding Hamiltonian for spinless fermions with NN and NNN interactions, $H=H_0+H_I$, with 
\begin{equation}
\label{tightbinding}
H_0=-t\sum_{\left\langle ij\right\rangle }(\mathit{e}^{\mathit{i}\theta_{ij}}c^{\dagger}_{i}c_{j}+ {\rm h.c.}),
\end{equation}
and 
\begin{equation}
\label{h1}
H_I=V_{1}\sum_{\left\langle ij\right\rangle }n_{i}n_{j}+V_{2}\sum_{\left\langle\langle ij \right\rangle\rangle }n_{i}n_{j}.
\end{equation}
Here $c^{\dagger}_{j}$ is the creation operator for a fermion on site $\br_j$ of the square or honeycomb lattice, the Peierls phase factor 
\begin{equation}
\theta_{ij}={2\pi\over\Phi_0}\int_{\br_i}^{\br_j}{\bA\cdot d{\bf l}}
\end{equation}
is defined on a link $\left\langle ij\right\rangle$, $t$ is the hopping amplitude between NN sites, $V_{1}$ and $V_{2}$  are the energy scales for NN and NNN interactions respectively and $n_{i}=c^{\dagger}_{i}c_{i}$ is the number operator. We focus on Hamiltonians $H_0$ that produce Dirac-type spectra for fermions in the absence of interactions.

In the following we treat $H_I$ in the mean-field (MF) approximation which should provide reliable information about the possible gapped phases in the phase diagram. We consider both the on-site and bond MF decoupling channels,
\begin{eqnarray}
\label{mean-field}
 n_{i}n_{j} &\rightarrow& n_{i}\langle n_{j} \rangle+n_{j}\langle n_{i} \rangle- \langle n_{i} \rangle\langle n_{j} \rangle \\
 n_{i}n_{j} &\rightarrow& -\Delta_{ij}c^{\dagger}_{i}c_{j}-\Delta_{ij}^{*}c^{\dagger}_{j}c_{i}+ \Delta_{ij}\Delta_{ij}^{*} 
\end{eqnarray}
where $\Delta_{ij}=\langle c^{\dagger}_{j}c_{i} \rangle$ with $i,j$ belonging to NN and NNN bonds.

\subsection{\label{pi flux}$\pi$-flux square lattice}
The system has a half magnetic flux quantum $\Phi_0=hc/e$ per plaquette. In this case the assumption of spinless fermions is quite natural because within the context of the realization described in Ref.\ \onlinecite{berciu1}  the electron spins would be polarized along the field.  As in Ref.\ \onlinecite{Ser08a}, we choose the Landau gauge $\vex A=(\Phi_0/2)(-y,0)$ where we have set the lattice spacing to unity and work at half filling. The unperturbed Hamiltonian then has a spectrum
\begin{equation}
E^{(0)}_\bk=\pm 2t\sqrt{\sin^2{k_x}+\cos^2{k_y}},
\end{equation}
with Dirac points at $(0,\pm \pi/2)$.

The repulsive interactions  can produce semi-metal (SM), charge density wave (CDW), quantum anomalous Hall (QAH), stripe and dimerized phases. The unit cell for this set up has four basis sites and is displayed in Fig.\ \ref{lattice}a. The SM phase corresponds to the undistorted $\pi$-flux lattice, which has no gap in the spectrum. 

The CDW corresponds to a modulation in charge density in both the $x$ and $y$ directions resulting in a checkerboard pattern on the lattice, whereas the stripe phase will be modulated only along one direction, hence giving rise to `stripes' of increased/decreased charge density. We use the following ansatz,  
\begin{equation}
\langle n_{i} \rangle = \frac{1}{2} + \rho(-1)^{i_{x}+i_{y}}+\nu(-1)^{i_{x}}
\end{equation}
for these two phases where $\br_i=(i_x,i_y)$.  

The QAH phase, which is characterized by a quantized Hall conductance without Landau levels,  occurs when a gap opens due to the spontaneous breaking of time reversal invariance. We require here that its order parameter, $\Delta_{ij}=\Delta_{\hat{x}+\hat{y}}$, produces a NNN hopping consistent with a half-flux quantum per unit cell. 

The dimerized phase, with its order parameter $\Delta_{ij}=\Delta_{\hat{x},\hat{y}}$  along NN this time, will act  to increase/decrease the existing hopping amplitude $t$. It can nominally take on any value locally, but we concentrate on either constant dimerization  or one that will support the quasi U(1) vortices mentioned above. We also note the energy scale $t$ becomes isotropically renormalized for any finite value of $V_{1}$ yielding a contribution $\delta t$.

Introducing the Fourier transform, $c_{j}={N}^{-1/2}\sum_{\bf k}e^{\mathit{i}\bf{k} \cdot \br_j}c_{\bf k}$, our Hamiltonian can then be brought into the matrix form in momentum space
\begin{equation}\label{hk}
H= \sum\limits_{\bf{k}}\Psi^{\dagger}_{\bf{k}}{\cal H}_{\bf{k}}\Psi_{\bf{k}} +E_{0}
\end{equation}
where $\Psi^{\dagger}_{\bf k} = \left(c^{\dagger}_{1\bf k},c^{\dagger}_{2\bf k},c^{\dagger}_{3\bf k},c^{\dagger}_{4\bf k}\right)$, and
\begin{align}
 E_{0}= N&\bigl[\frac{\bar{\rho}^{2}}{8\left(V_{1}-V_{2}-V_{3}\right)}+\frac{\bar{\nu}^{2}}{8\left(V_{2}-V_{3}\right)} + \frac{\eta_{x}^{2}+\eta_{y}^{2}}{V_{1}}\nonumber\\
 &+\frac{2}{V_{2}}\xi^{2}+\frac{2}{V_{1}}\delta t^{2}\bigr]
\end{align}
with
 $\bar{\rho}=4\rho(V_{2}+V_{3}-V_{1})$, $\bar{\nu}=4\nu(V_{3}-V_{2})$,
$\eta_{x}=V_{1}\Delta_{\hat{x}}$, $\eta_{y}=V_{1}\Delta_{\hat{y}}$ and $\xi = \Delta_{\hat{x}+\hat{y}}V_{2}$.
The Hamiltonian matrix reads
\begin{equation}
{\cal H}_{\bk} = \left(
	\begin{matrix}
	\bar{\rho}+\bar{\nu}&\Gamma&\Omega_{x}&-\Omega_{y}^{*}\\
        \Gamma^{\ast}&\bar{\rho}+\bar{\nu}&-\Omega_{y}&-\Omega_{x}^{*}\\
	\Omega_{x}^{\ast}&-\Omega_{y}^{*}& -\bar{\rho}-\bar{\nu}&\Gamma\\
        -\Omega_{y}&-\Omega_{x}&\Gamma^{\ast}& -\bar{\rho}-\bar{\nu}
	\end{matrix}\right)
\end{equation}
with
\begin{eqnarray}
 \Omega_{x,y}&=& 2(\tilde{t}\cos k_{x,y}+\mathit{i}\eta_{x,y}\sin k_{x,y}),\\ \Gamma &=& 2\mathit{i}\xi[\cos(k_{x}+k_{y})-\cos(k_{y}-k_{x})]
\end{eqnarray}
and $\tilde{t}=t+\delta t$. For reasons that will become apparent shortly we have also included in the above analysis the third-nearest-neighbor (NNNN) repulsion $V_3$.

We can now diagonalize ${\cal H}_{\bk}$ and obtain the exact dispersion (setting the chemical potential to zero) with four particle-hole symmetric branches
\begin{align}
&E_{\bf k}  = \pm\bigl[\bar{\nu}^{2}+\bar{\rho}^{2}+4\tilde{t}^{2}\left(\mathit{c}^{2}_{x} +\mathit{c}^{2}_{y}\right)
+4\left(\eta_{x}^{2}\mathit{s}^{2}_{x}+\eta_{y}^{2}\mathit{s}^{2}_{y}\right)\nonumber\\
&+16\xi^{2}\mathit{s}^{2}_{x}\mathit{s}^{2}_{y}\pm2\bigl(\bar{\nu}^{2}\bar{\rho}^{2}+4\bar{\nu}^{2}\tilde{t}^{2}\mathit{c}^{2}_{y}
-32\bar{\nu}\tilde{t}\eta_{x}\xi\mathit{c}_{y}\mathit{s}^{2}_{x}\mathit{s}_{y}\\
&+4\bar{\nu}^{2}\eta_{y}^{2}\mathit{s}^{2}_{y}+16\bar{\rho}^{2}\xi^{2}\mathit{s}^{2}_{x}\mathit{s}^{2}_{y}+64\xi^{2}\mathit{s}^{2}_{x}\mathit{s}^{2}_{y}(\eta_{x}^{2}\mathit{s}^{2}_{x}+\eta_{y}^{2}\mathit{s}^{2}_{y})\bigr)^\frac{1}{2}\bigr]^\frac{1}{2}.\nonumber
\end{align}
Here $\bf k$ is taken over the reduced Brillouin zone $-\frac{\pi}{2}\leq k_{x}\leq\frac{\pi}{2}$, $-\frac{\pi}{2}\leq k_{y}\leq\frac{\pi}{2}$ and $c_{x}=\cos k_{x}$, $s_{y}=\sin k_{y}$
, etc.

\begin{figure}
\includegraphics[width=7cm]{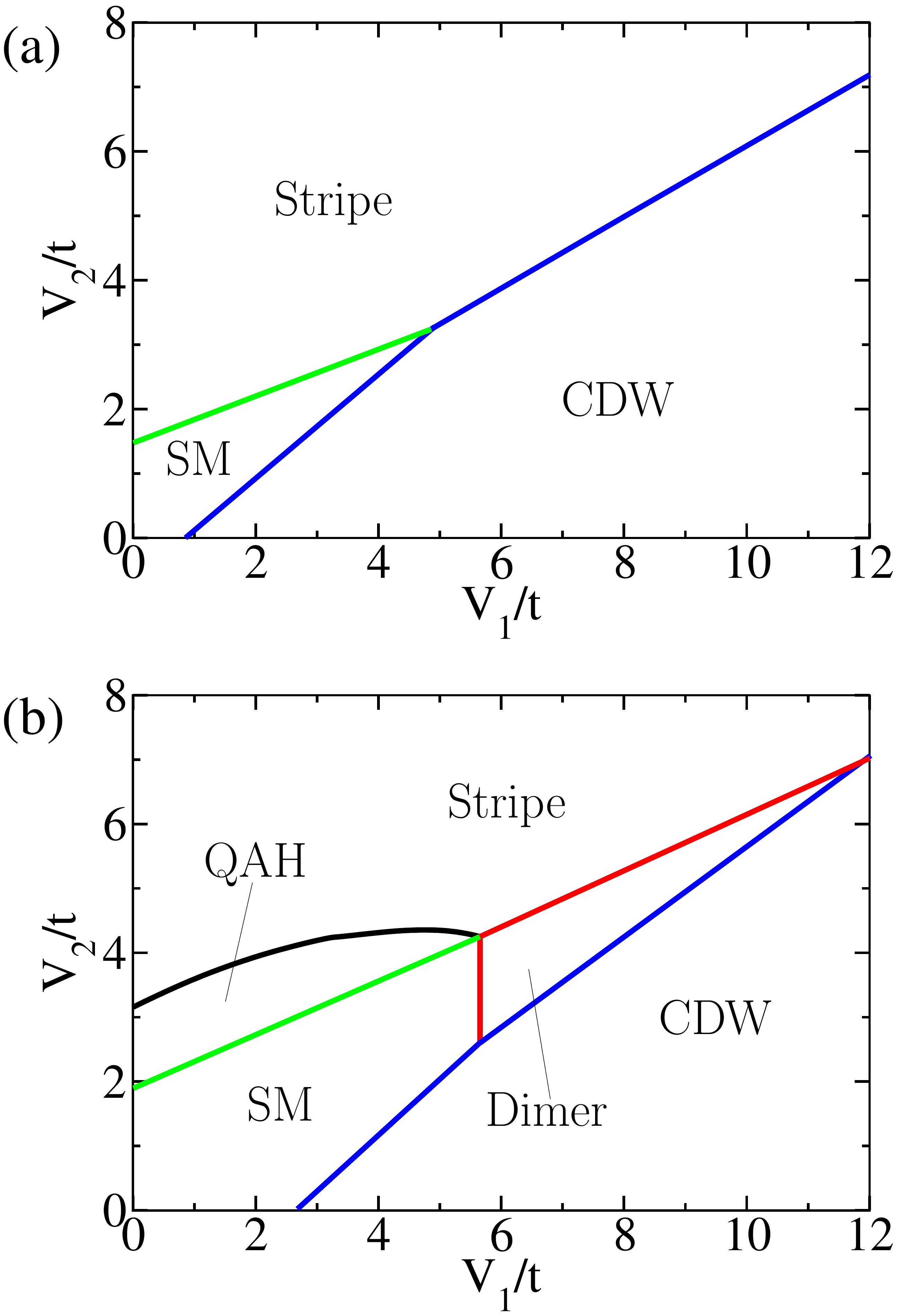}
\caption{Phase diagram for $\pi$-flux model with spinless fermions. (a) $V_{3}=0$. (b) $V_{3}=1.5t$  } 
\label{pda}
\end{figure}

From here,  we may calculate the free energy 
\begin{equation}\label{free}
F=-\frac{1}{\beta}\sum_{\vex k}\ln(1+\mathit{e}^{\beta E_{\vex k}}) +E_{0}
\end{equation}
where $\beta = 1/\mathit{k}_B T$ and the summation is over all four branches of $E_\bk$. The ground state is then obtained by minimizing $F$ with respect to each of the order parameters. This yields a set of gap equations (listed in Appendix A) from which the phase diagram (at $T=0$) seen in Fig.\ \ref{pda}a below follows after a numerical self-consistent iteration. The transition from the SM phase to the  CDW is second order, whereas all other transitions are first order. 

We observe that, within the present model with $V_1$ and $V_2$, the stripe instability completely wipes out the QAH and the dimerized phases expected to occur at finite $V_2$.  In order to suppress the stripe phase we include a further term in the Hamiltonian above, namely a $V_{3}$ term for NNNN interactions. Such term will generically be present in any system that can support NN and NNN interactions. To simplify matters we shall ignore any contribution to the kinetic energy generated  by $V_3$ and focus on the frustration it generates for the stripe instability. We find that for $V_3$ non-zero  both the QAH and the dimerized phases are stabilized (Fig.\ \ref{pda}b) although stripes reappear at stronger $V_2$.

\subsection{\label{honeycomb}Honeycomb lattice}
On the honeycomb lattice we may set $\theta_{ij}=0$ (no magnetic field needed for Dirac band structure). The unperturbed spectrum reads
\begin{equation}
E^{(0)}_\bk=\pm t|\tau(\bk)|
\end{equation}
where $\tau\left({\bf k}\right)=\sum_{i}e^{\mathit{i}\bk \cdot {\bf a}_{i}}$, with vectors ${\bf a}_{i}$ shown in Fig.\ \ref{lattice}b.
Also, there is no option of including a stripe phase, leaving us with the CDW, Kekul\'e and QAH phases.
As seen in Fig.\ \ref{lattice}b the system now has 6 atoms in the unit cell.

For the CDW we set 
\begin{equation}
\rho=\frac{1}{2}\left(\langle n_{i}^A \rangle-\langle n_{i}^B \rangle\right),
\end{equation}
where A and B refer to the two sublattices of the undistorted honeycomb lattice, which leads to the contribution (each site having 3 NN and 6 NNN sites)
\begin{equation}
\delta H_{\rm CDW}=3\rho\left(V_{1}-2V_{2}\right)\sum\limits_{i}n_{i}\left(-1\right)^{\mathit{i}}+\frac{3}{2}N\rho^{2}\left(V_{1}-2V_{2}\right)
\end{equation}

For the Kekul\'e phase, we again use the decoupling in equation (\ref{mean-field}) where $c^{\dagger}_{i}c_{j}\langle c^{\dagger}_{j}c_{i} \rangle$ simply adds to the  $t c^{\dagger}_{i}c_{j}$ term in $H_0$ according to
 \begin{equation}
 t \rightarrow t_{a}=t+\delta t+V_{1}\eta_{a}, \quad a=1,2,3
 \end{equation}
and
 \begin{equation}
 \sum\limits_{ij}V_{ij}\langle c^{\dagger}_{i}c_{j} \rangle\langle c^{\dagger}_{j}c_{i} \rangle=V_{1}\left(\eta_{1}^{2}+\eta_{2}^{2}+\eta_{3}^{2}\right)\frac{N}{2}
 \end{equation}
The index $a$ above labels the three bonds emanating from each A site. We may also write
\begin{equation}
\eta_{a}=\eta\cos\left(\varphi+\frac{2\pi}{3}a\right),
\label{param}
\end{equation}
where $\varphi$ parametrizes the most general Kekul\'e distortion. We shall find below that the ground state energy is minimized by $\varphi=0,2\pi/3,4\pi/3$, corresponding to the three degenerate Kekul\'e ground states. For any $\varphi$ it holds that $\sum\limits_{a}\eta_{a}=0$, and $\sum\limits_{a}\eta_{a}^{2}=\frac{3}{2}\eta^{2}$.

For the QAH phase we have $\langle c^{\dagger}_{i}c_{j} \rangle\equiv\xi_{ij}=\pm\mathit{i}\xi_{\alpha}$ (for $i, j\in$ NNN)  where $\alpha=A,B$, and the sign is taken according to the arrow in Fig.\ \ref{lattice}b. This generates a contribution
\begin{equation}
\delta H_{\rm QAH}=V_2\sum_{\langle\langle ij \rangle\rangle}(\xi_{ij}c^\dagger_ic_j+{\rm h.c.})+3NV_2(\xi_A^2+\xi_B^2)
\end{equation}
to the mean-field Hamiltonian.

Combining these contributions and Fourier transforming we arrive at the $k$-space Hamiltonian of the form Eq.\ (\ref{hk}) with $\Psi_{\bk}$ a six-component spinor and
\begin{equation}\label{hh}
{\cal H}_{\bf{k}} = 
 \left( \begin{array}{cccccc}
\bar{\rho} & \bar{\xi_{A}}\mathit{S}_{\bk} & \bar{\xi_{A}}\mathit{S}_{\bk}^{*} & t_{3}\mathit{e}^{\mathit{i}k_{1}}&t_{2}\mathit{e}^{\mathit{i}k_{3}} & t_{1}\mathit{e}^{\mathit{i}k_{2}} \\
 & \bar{\rho} & \bar{\xi_{A}}\mathit{S}_{\bk} & t_{2}\mathit{e}^{\mathit{i}k_{2}} & t_{1}\mathit{e}^{\mathit{i}k_{1}} & t_{3}\mathit{e}^{\mathit{i}k_{3}} \\
 &   & \bar{\rho} & t_{1}\mathit{e}^{\mathit{i}k_{3}} & t_{3}\mathit{e}^{\mathit{i}k_{2}} & t_{2}\mathit{e}^{\mathit{i}k_{1}} \\
 &   &   &  -\bar{\rho}  & -\bar{\xi_{B}}\mathit{S}_{\bk} &  -\bar{\xi_{B}}\mathit{S}_{\bk}^{*} \\
 &   &   &   & -\bar{\rho} & -\bar{\xi_{B}}\mathit{S}_{\bk}\\
 &   &   &  &  & -\bar{\rho} 
\end{array} \right).
\end{equation}
Here $\bar{\xi_{\alpha}}=\xi_{\alpha}V_{2}$, $\bar{\rho}=3\rho\left(V_{1}-2V_{2}\right)$, $k_i={\bf a}_i\cdot \bk$,
\begin{equation}
\mathit{S}_{k}=-\mathit{i}\left[\mathit{e}^{\mathit{i}\left(k_{2}-k_{3}\right)}
+\mathit{e}^{\mathit{i}\left(k_{3}-k_{1}\right)}
+\mathit{e}^{\mathit{i}\left(k_{1}-k_{2}\right)}\right]
\end{equation}
and
\begin{equation}
E_{0}=\frac{N}{2}\left[\frac{\bar{\rho}^{2}}{3\left(V_{1}-2V_{2}\right)}+\frac{\bar{\eta}^{2}}{V_{1}} +3\frac{\bar{\xi_{A}}^{2}+\bar{\xi_{B}}^{2}}{V_{2}}+3\frac{{\delta t}^{2}}{V_{1}}\right].
\end{equation}
The reduced Brillouin zone, which is derived below in the Appendix B, is comprised of the following set of points
\begin{equation}
\left(k_{1},k_{2},k_{3}\right)=\frac{2\pi}{3}\frac{1}{L}\left(n-m,m,-n\right)\quad 
\end{equation}
where $m,n=1,2,....,L$ and $6L^{2}=N$ is the total number of sites.

It is possible to find exact eigenvalues of the $6\times 6$ matrix indicated in Eq.\ (\ref{hh}) by noticing that ${\cal H}_\bk^2$ is block-diagonal with two $3\times 3$ blocks on the diagonal. Unfortunately, the eigenvalues obtained as roots of the associated cubic secular equations are given by lengthy and complicated expressions, which do not lend themselves to a convenient analysis. For this reason we choose to calculate the eigenvalues of ${\cal H}_\bk$ numerically for a dense discrete set of momenta $\bk$ in the first BZ. We use these eigenvalues to compute the free energy, Eq.\ (\ref{free}), for a set of mean field parameters $\bar{\rho}$, $\bar{\eta}$, $\varphi$ and $\bar{\xi}_\alpha$ and given fixed values of $V_1$ and $V_2$. The ground state of the system is then found by minimizing the free energy (at $T=0$) with respect to the above MF parameters. This is achieved by employing the standard Powel multivariate minimization routine.
\begin{figure}
\includegraphics[width=6cm]{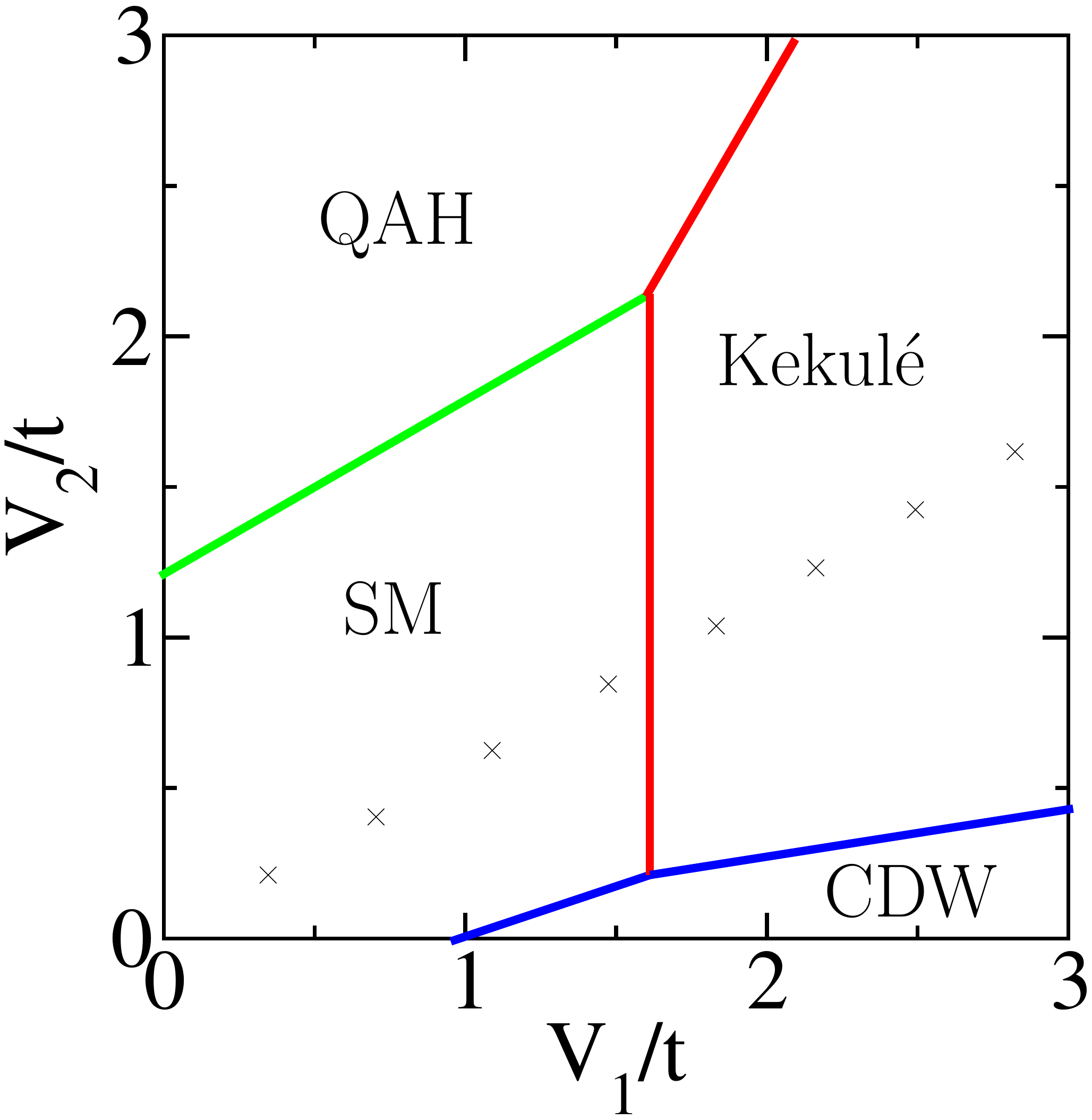}
\caption{Phase diagram for honeycomb lattice. At the mean-field level all transitions from the SM phase are second order whereas transitions between all the gapped phases are first order. The $\times$'s represent the line along which the ratio $V_{2}/V_{1}$ would be expected to fall in graphene based on a crude estimate of the bare Coulomb repulsion.\cite{drut1}} 
\label{pdc}
\end{figure}

The resulting phase diagram is displayed in Fig.\ \ref{pdc}. We observe that in addition to the CDW and QAH phases identified previously\cite{Raghu} a large portion of the phase diagram is occupied by the Kekul\'e phase.

Although the critical lines in the phase diagram are determined numerically,
the critical couplings for the CDW at $V_2=0$ and for QAH at $V_1=0$ can be found exactly from the corresponding gap equations for the model excluding Kekul\'e order. These read
\begin{eqnarray}
\frac{\tilde{t}}{V_{1c}}&=&\frac{3}{N}\sum_{{\bf k}}\frac{1}{|\tau\left({\bf k}\right)|}\simeq 1.341,
\\
\frac{\tilde{t}}{V_{2c}}&=&\frac{2}{3N}\sum_{{\bf k}}\frac{\left(\sum_{i}\sin {\bf k}\cdot {\bf b}_{i}\right)^{2}}{|\tau\left({\bf k}\right)|}\simeq 0.840,
\end{eqnarray}
where the ${\bf b}_{i}$ correspond to the set of NNN vectors for either of the triangular sublattices inside a single plaquette. If we wish to express critical couplings in terms of the bare hopping amplitude $t$ we must include the MF equation for the hopping renormalization
\begin{equation}
\delta t=V_{1}\frac{1}{3N}\sum_{\bf k}|\tau\left({\bf k}\right)|
\simeq 0.262 V_{1}
\end{equation}
We thus obtain $V_{1c}\simeq 0.93t$ and $V_{2c}\simeq 1.20t$, in good agreement with the numerical results of Fig.\ \ref{pdc}. Our value of $V_{2c}$ also agrees with Ref.\ \onlinecite{Raghu}, but our $V_{1c}$ is smaller by a factor of about 1.5 when expressed in terms of $t$  and about a factor of 2 in terms $\tilde{t}$. We do not know what is the reason for this discrepancy. Since the critical couplings quoted above agree with our numerically determined phase diagram, we are confident that these are correct within the definitions employed in this study.

\section{\label{vortex}Self-consistent vortex structure}
In a superconductor or a superfluid the order parameter has a global U(1) symmetry related to its complex phase. This means that, even on a lattice, a U(1) vortex is a legitimate topological defect. In the Kekul\'e (dimerized) phase the order parameter exhibits a global Z$_3$ (Z$_4$) symmetry leading to the possibility of a  Z$_3$ (Z$_4$) vortex, which can be pictured as a point where the corresponding 3 (4) domains meet. Since the domain walls cost an energy per unit length, the energy of a single isolated Z$_n$ vortex diverges linearly with the system size and, equivalently, a vortex-antivortex pair experiences linear confinement. U(1) vortices, on the other hand, experience only logarithmic confinement. This has important consequences for the appearance of vortices in these systems. U(1) vortices can thermally unbind above the Kosterlitz-Thouless transition temperature $T_{KT}$ while Z$_n$ vortices remain confined at all temperatures. 

It has been suggested\cite{Ser08a} that the Z$_n$ vortices in the Kekul\'e/dimerized phase can nevertheless be observed since at short length scales they resemble U(1) vortices, and thus, for relatively short  intervortex separations, interact only logarithmically. This has to do with the fact that the energy cost of a domain wall is relatively small. At length scales exceeding the confinement length $\zeta_{\rm conf}$, vortices remain linearly confined, as dictated by symmetry. If $\zeta_{\rm conf}$ is sufficiently long compared to the vortex core size $\zeta_v$, however, then the zero mode and the fractional charge associated with an individual vortex could be observed experimentally. To address this issue quantitatively we now carry out a fully self-consistent calculation of a vortex in the dimerized phase on the $\pi$-flux lattice. Our calculations show that, indeed, at short length scales, a vortex in the dimer order parameter resembles a U(1) vortex while on longer length scales clear domains separated by domain walls emerge.

Within our region of parameter space on the square lattice having a stable dimerized phase, we set up a discretized version of a U(1) vortex in the dimerization pattern $\Delta_{ij}$ as described in Ref.\ \onlinecite{Ser08a}.
For this initial vortex the corresponding MF Hamiltonian
\begin{equation}
\label{hmf}
H_{\rm MF}=H_0+\sum_{\left\langle ij\right\rangle} (\Delta_{ij}c^\dagger_{i}c_{j}+{\rm h.c.})
\end{equation}
is diagonalized and the new order parameter is found using the condition
\begin{equation}\label{self}
\Delta_{ij}=\langle c_{i}^{\dagger}c_{j}\rangle=
U_{li}^{\dagger}U_{j k}\langle d_{l}^{\dagger}d_{k}\rangle
= \sum_{l}U_{li}^{\dagger}U_{jl}\mathit{f}(\epsilon_{l})
\end{equation}
where $\mathit{f}$ is the fermi function, and $\epsilon_l$ and $d_j$ are the eigenvalues and  the eigenmodes of $H_{\rm MF}$. The unitary matrices $U_{ij}$ are comprised of the eigenvectors of the corresponding Hamiltonian matrix $\mathcal{H}_{ij}$, which we find numerically through exact diagonalization.
 Eqs.\ (\ref{hmf},\ref{self}) are then iterated to self-consistency. We have solved the system for odd lattice sizes up to $49\times 49$ using open boundary conditions with a vortex positioned at the central site.

Our results for the self-consistent vortex structure are summarized in Fig.\ \ref{vortex}. Instead of (real) bond fields $\Delta_{ij}$ it is useful to consider a complex on-site `dimer' order parameter defined as
\begin{equation}\label{gi}
g_i = e^{i{\pi \over 4}}\sum_{\hat\mu}\gamma_{i,i+\hat\mu} \Delta_{i,i+\hat\mu}.
\end{equation}
Here the $\hat\mu$ are the four nearest-neighbor unit vectors, 
$\gamma_{i,i+\hat x}=(-1)^{i_x}$, $\gamma_{i,i+\hat y}=\i(-1)^{i_y}$ and 
$\gamma_{i,i-\hat\mu}=\gamma_{i-\hat\mu,i}$. The phase of $g_i$ contains information on the local orientation of the dimer pattern, e.g.\ $g_i\sim e^{i{\pi\over 2}(n-{1\over 2})}$ with $n=1,2,3,4$ describes four basic `columnar' patterns modulated along the $x$ and $y$ directions while  $g_i\sim e^{i{\pi\over 2}n}$ describes four possible `box' phases with the strength of dimerization modulated along both the $x$ and $y$ directions. We note that in the uniform system the box phases correspond to the ground states.

The vortex has a finite size core with a radius inversely proportional to the gap $\Delta$, as seen in Fig.\ \ref{vortex}b. Near the core the phase behaves as in a U(1) vortex although at longer distances domain walls are seen to form along the lattice diagonals, Fig.\ \ref{vortex}a, characteristic of a Z$_4$ vortex. A net fractional charge $e/2$ accumulates near the core, Fig.\ \ref{vortex}c, as expected on the basis of general arguments.\cite{HouChaMud07a,Ser08a}

The nature of the dimerized box ground state and the competition with other possible states can be understood from the following analysis.
\begin{figure}
\includegraphics[width=3.3in]{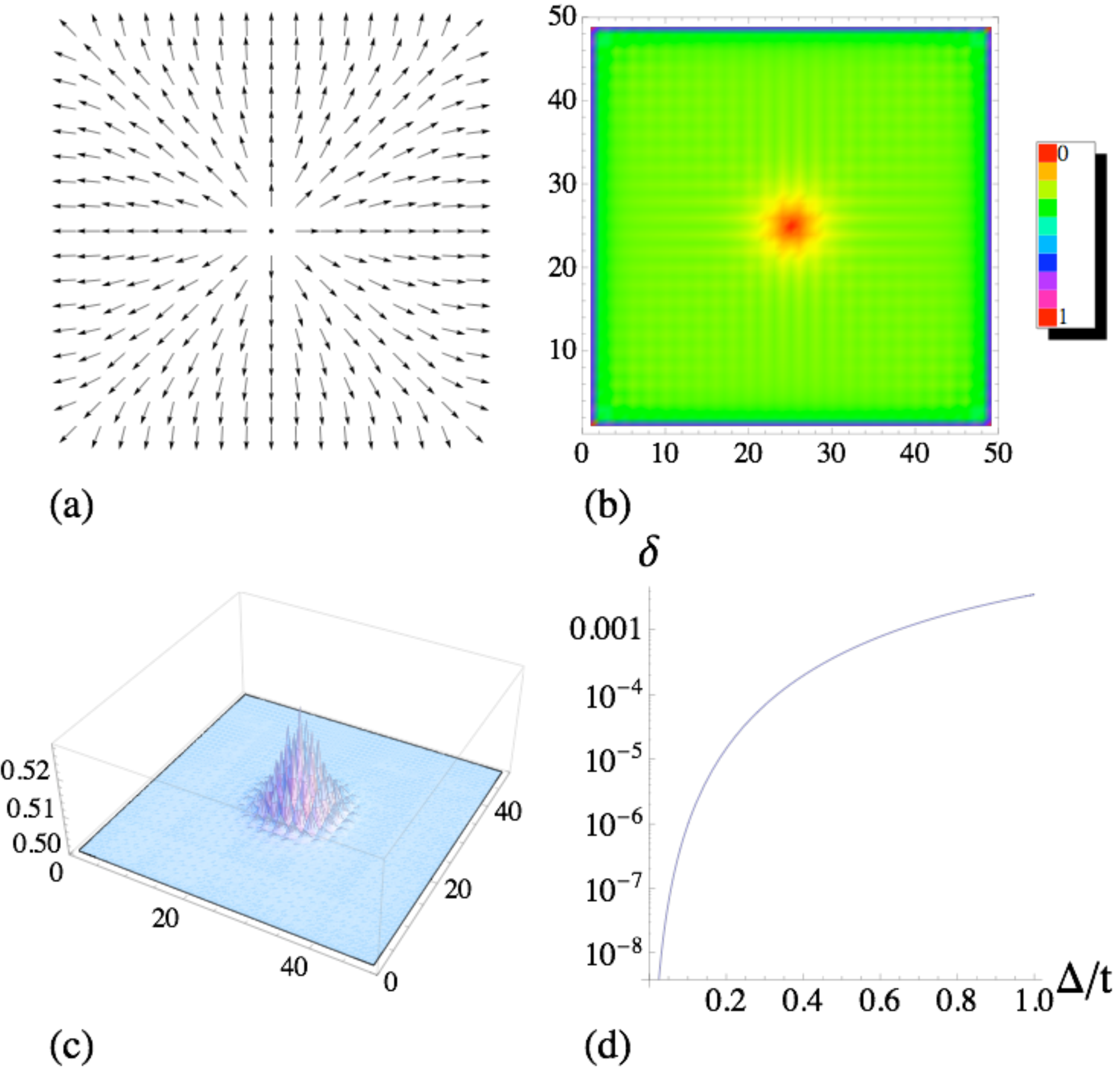}
\caption{Self-consistent vortex structure in the dimerized phase.  (a) Phase  and (b) magnitude of the on-site dimer order parameter $g_i$ defined in Eq.\
(\ref{gi}). (c) Charge density profile showing excess charge density in the vortex core. The total accumulated charge is $e/2$.  (d) Anisotropy parameter $\delta$ defined in Eq.\ (\ref{delta}). For figures (a)-(c), $\Delta/t\approx 0.24$} \label{vortex}
\end{figure}
If we tune $V_{1}$ and $V_{2}$ so that we are in the dimerized phase, then all other order parameters can be set to zero. In the uniform state we can
parameterize $\eta_{x}$ and $\eta_{y}$ by an angle $\chi$, so that ${\bf\Delta}=(\eta_{x},\eta_{y})=\Delta(\cos\chi,\sin\chi)$, and write our ground state energy per site as (summing over all negative energy states)
\begin{equation}
E_{0}(\chi)=-{\sum_{\bf k}}'\sqrt{\epsilon_\bk^2+\omega_\bk^2}
\end{equation}
with 
\begin{equation}
\begin{split}
\epsilon^2_\bk&=4t^2(\cos^{2}k_{x} + \cos^{2}k_{y})
+\frac{\Delta^{2}}{2}(\sin^{2}k_{x} + \sin^{2} k_{y}),\nonumber\\
\omega^2_\bk&=\frac{\Delta^{2}}{2}\cos 2\chi (\sin^{2}k_{x} - \sin^{2} k_{y}).
\end{split}
\end{equation}
Taylor expanding 
\begin{equation}
\epsilon_{\bf k}\sqrt{1+\frac{\omega^{2}_{\bf k}}{\epsilon^{2}_{\bf k}}}\simeq\epsilon_{\bf k}\left(1+\frac{1}{2}\frac{\omega^{2}_{\bf k}}{\epsilon^{2}_{\bf k}}-\frac{1}{8}\frac{\omega^{4}_{\bf k}}{\epsilon^{4}_{\bf k}}\right)
\end{equation}
we then have to fourth order in $\omega_{\bf k}$
\begin{equation}
\label{anisotropy}
E_{0}\left(\chi\right)=-\sum_{\bf k}\epsilon_{\bf k}+\frac{\Delta^{4}}{32}\cos^{2}2\chi\sum_{\bf k}\frac{(\sin^{2}k_{x}- \sin^{2} k_{y})^{2}}{\epsilon^{3}_{\bf k}}.
\end{equation}
The ground state is clearly minimized by $\chi=\pm\frac{n\pi}{4}$ for $n=1,3$, corresponding to the box phase. The columnar phase corresponds to the maximum of energy. However, by plotting the anisotropy ratio 
\begin{equation}\label{delta}
\delta={E_0(0)-E_0(\pi/4)\over E_0(\pi/4)}
\end{equation}
against $\Delta/t$, one can see, as in  Fig.\ \ref{vortex}d, that the difference between the maximum and minimum values of $E_{0}(\chi)$ will be very small for a broad range of values of $\Delta/t$. The smallness of $\delta$ explains why at short length scales our Z$_4$ vortex behaves as a U(1) vortex.

\section{\label{spin}Spinful Fermions}

For spinful Fermions a Hubbard term $U\sum_in_{i\uparrow}n_{i\downarrow}$ must be added to the interaction Hamiltonian, Eq.\ (\ref{h1}), reflecting the strong on-site Coulomb repulsion between electrons of opposite spin. It is well-known that for a bi-partite lattice, a large $U$ favors the antiferromagnetic SDW state. The mean-field phase diagram for spinful electrons on the honeycomb lattice in the $U$-$V_1$-$V_2$ space has been mapped out in Ref.\ \onlinecite{Raghu}. For small values of $U$ the phase diagram resembles that shown in Fig.\ \ref{pdc} except that the QAH phase is replaced by the QSH phase. The two are degenerate at the mean field level but quantum fluctuations favor the latter.

A question that we would like to answer is how the $U$-$V_1$-$V_2$ phase diagram on the honeycomb lattice is affected by the presence of the Kekul\'e phase, which was not considered in  Ref.\ \onlinecite{Raghu}. Is there a region in the parameter space where a `spin Kekul\'e' phase is stabilized? The latter represents a spinful version of the ordinary Kekul\'e phase and its order parameter is defined as follows. The nearest neighbor interactions in $H_I$ can be re-written using the identity $ (n_{i } -1)(n_{j}-1) =1  -\frac{1}{2} \left( \chi^{\mu }_{ij} \right)^{\dagger}\chi^{\mu}_{ij} $, where $\chi^{\mu}_{ij} = c^{\dagger}_{i \alpha} \sigma^{\mu}_{\alpha
\beta} c_{j \beta}$, $\mu = 0 \ldots 3$ and $\sigma^{\mu} =\left( 1, \bm \sigma \right)$. In a mean field picture, then,  $\langle \chi^0 \rangle \neq 0$ corresponds to the standard Kekul\'e phase and $\langle \chi^{i} \rangle \neq 0$ describes the spinful version. 

We find, however, that the spin Kekul\'e phase is not a mean field
ground state of our Hamiltonian. Instead, a state in which two
projections of the spin form the ordinary Kekul\'e phase is favored.
This can be seen from the following symmetry based argument. We focus
on the $\mu=3$ state and put the spin up into the ordinary Kekul\'e
phase with parameters $(\eta,\varphi)$ that minimize its ground state
energy.  The spin down electrons must then go into the Kekul\'e phase
with parameters $(-\eta,\varphi)$ due to the $\sigma_{3}$ in the order
parameter. However, this is not a ground state for spin down
electrons, meaning that the overall energy must be higher than the
$\mu=0$ state. Thus the energy of the SK phase is not exactly
degenerate with the standard Kekul\'e phase as is the case for the
QAH/QSH spectrum, but instead has a small splitting with the Kekul\'e
phase being favored as the ground state.  So, in absence of some type
of interaction that could favor the SK phases, our current Hamiltonian
is not capable of producing this potentially interesting phase.
\newline

\section{Conclusions}

Our studies show that interacting fermions moving on the honeycomb and
$\pi$-flux square lattices can form a stable Kekul\'e phase, in
addition to the previously identified CDW, SDW, and QAH/QSH
phases. Two questions naturally arise: Can the Kekul\'e phase be
stabilized in a realistic system and if so, how can it be
experimentally observed? Inspection of the phase diagram shown in
Fig.\ 2 reveals that the prospects of observing the Kekul\'e
(dimerized) phase on the square lattice are not very good. When the
NNN interaction is sufficiently strong to suppress the CDW, one obtains
the stripe phase. Even when this is suppressed by introduction of NNNN
repulsion, the dimerized phase appears only at strong coupling
($V_1\simeq 6t$) and only in a small sliver of the parameter space. We
thus conclude that although the square lattice is very convenient from
the point of view of theoretical considerations, it is an unlikely
candidate for the experimental observation of the Kekul\'e/dimerized
phase.

Prospects for the Kekul\'e phase appear much better on the honeycomb
lattice as it occupies a large portion of the $V_1$-$V_2$ mean field
phase diagram shown in Fig.\ 3. The Kekul\'e phase should appear
whenever the interaction strength becomes large and $V_1$, $V_2$ are
comparable. In natural graphene the Coulomb interaction is strong but
evidently not strong enough to open up a gap at the Dirac
point,\cite{graphene} at least when a sample is placed on a substrate
such as SiO$_2$. It has been suggested recently, on the basis of large
scale Monte Carlo simulations, that graphene freely suspended {\em in
vacuum} could become an insulator when the Coulomb interaction between
electrons is accounted for.\cite{drut1} The difference between the
substrate and vacuum arises from the dielectric constant, which weakens
the Coulomb interaction in the vicinity of a polarizable solid such
as SiO$_2$, and also because of disorder induced by the
substrate. Although Ref.\ \onlinecite{drut1} focused on the CDW
instability, one may argue based on our present mean field study that
the Kekul\'e phase is a more likely candidate for an instability
induced by the Coulomb interaction. In a very crude estimate the ratio
$V_2/V_1$ should be equal to the ratio between the NN and NNN bond
lengths, i.e.\ $V_2/V_1\approx 1/\sqrt{3}\simeq 0.5773$. As
illustrated in Fig.\ 3, this ratio of $V_2/V_1$ clearly favors the
Kekul\'e phase over CDW if the overall effective interaction strength
is strong enough to open up a gap. Recent experimental observation of
the fractional quantum Hall effect in suspended graphene\cite{du1}
confirms the increased importance of interactions over graphene on
a substrate, where the fractional quantum Hall effect thus far eluded
experimental detection. At zero field, however, even the suspended
samples appear semi-metallic\cite{du1} down to 1.2K, suggesting that
Coulomb interactions are still too weak to open up a significant gap.
We note, finally, that the Kekul\'e phase has been proposed to emerge
as the leading instability of graphene in an applied magnetic
field.\cite{Hou2} Also, a phase analogous to the Kekul\'e phase in
graphene has been proposed to exist for electrons on the kagome
lattice at 1/3 filling.\cite{guo1}

The Kekul\'e phase has broken translational symmetry characterized by
a wavevector connecting the two inequivalent Dirac points in the
graphene spectrum. The on-site charge density remains uniform and this
limits the spectrum of experimental probes capable of detecting this
pattern of symmetry breaking. Here we propose to use the technique of
Fourier transform scanning tunneling spectroscopy (FT-STS) that has
already been applied to graphene.\cite{mallet1} With sufficient
resolution, FT-STS allows the mapping out of fine details of quasiparticle
interference patterns at non-zero momenta, and, with help from the
theoretical modeling of such patterns,\cite{tami1} it should be
possible to establish the existence of the Kekul\'e or other symmetry
breaking phases.

\emph{Acknowledgment.}---The authors have benefited from discussions with B. Seradjeh and G. Semenoff. This work was supported by NSERC, CIfAR and the Aspen Center of Physics.

~

\section{Appendix}

\subsection{Gap Equations}
For completeness we list the gap equations for the $\pi$-flux model on the square lattice in the limit $T\rightarrow 0$ for $\bar{\nu}$,  $\bar{\rho}$, $\eta_{x}$, $\eta_{y}$ , $\xi$ and $\delta t$ :

\begin{widetext}
\begin{align}
\bar{\rho}&=\frac{4\left(V_{1}-V_{2}-V_{3}\right)}{N}\sum_{{\bf k},\pm}\frac{1}{\left|E_{k,\pm}\right|}\bigl[\bar{\rho}\pm\frac{1}{\tilde{E}_{k}}\bigl(\bar{\nu}^{2}\bar{\rho}+16\bar{\rho}\xi^{2}\sin^{2}k_{x}\sin^{2}k_{y}\bigr)\bigr]\\
\bar{\nu}&=\frac{4\left(V_{2}-V_{3}\right)}{N}\sum_{k,\pm}\frac{1}{\left|E_{k,\pm}\right|}\bigl[\bar{\nu}\pm\frac{1}{\tilde{E}_{k}}\bigl(\bar{\nu}\bar{\rho}^{2}+4\bar{\nu}\tilde{t}^{2}\cos^{2} k_{y}-16\eta_{x}\xi\cos k_{y}\sin^{2}k_{x}\sin k_{y}+4\bar{\nu}\eta_{y}^{2}\sin^{2}k_{y}\bigr)\bigr]\\
\eta_{x}&=\frac{2V_{1}}{N}\sum_{k,\pm}\frac{1}{\left|E_{k,\pm}\right|}\bigl[\eta_{x}\sin^{2}k_{x}\pm\frac{1}{\tilde{E}_{k}}\sin^{2}k_{x}\bigl(16\xi^{2}\eta_{x}\sin^{2}k_{x}\sin^{2}k_{y}-4\bar{\nu}\tilde{t}\xi\cos k_{y}\sin k_{y}\bigr)\bigr]\\
\eta_{y}&=\frac{2V_{1}}{N}\sum_{k,\pm}\frac{1}{\left|E_{k,\pm}\right|}\bigl[\eta_{y}\sin^{2}k_{y}\pm\frac{1}{\tilde{E}_{k}}\bigl(\bar{\nu}^{2}\eta_{y}\sin^{2}k_{y}+16\xi^{2}\eta_{y}\sin^{4}k_{y}\sin^{2}k_{x}\bigr)\bigr]\\
\xi&=\frac{4V_{2}}{N}\sum_{k,\pm}\frac{\sin^{2}k_{x}}{\left|E_{k,\pm}\right|}\bigl[\xi\sin^{2}k_{y}\pm\frac{1}{\tilde{E}_{k}}\bigl(\bar{\rho}^{2}\xi\sin^{2}k_{y}+4\xi (\eta_{x}^{2}\sin^{2}k_{x}\sin^{2}k_{y}+\eta_{y}^{2}\sin^{4}k_{y})-\bar{\nu}\tilde{t}\eta_{x}\cos k_{y}\sin k_{y}\bigr)\bigr]\\
\delta t &=\frac{V_{1}}{N}\sum_{k,\pm}\frac{1}{\left|E_{k,\pm}\right|}\bigl[\tilde{t}\left(\cos^{2} k_{y}+\cos^{2} k_{y}\right)\pm\frac{1}{\tilde{E}_{k}}\bigl(2\bar{\nu}^{2}\tilde{t}\cos^{2} k_{y}-8\bar{\nu}\tilde{t}\eta_{x}\xi\cos k_{y}\sin^{2}k_{x}\sin k_{y}\bigr)\bigr]\\ \nonumber
\end{align} 
where
\begin{align}
\tilde{E}_{k}=\bigl(\bar{\nu}^{2}\bar{\rho}^{2}&+4\bar{\nu}^{2}\tilde{t}^{2}\cos^{2} k_{y}
-32\bar{\nu}\tilde{t}\eta_{x}\xi\cos k_{y}\sin^{2}k_{x}\sin k_{y}+4\bar{\nu}^{2}\eta_{y}^{2}\sin^{2}k_{y}\nonumber \\ 
&+16\bar{\rho}^{2}\xi^{2}\sin^{2}k_{x}\sin^{2}k_{y}+64\xi^{2}\sin^{2}k_{x}\sin^{2}k_{y}(\eta_{x}^{2}\sin^{2}k_{x}+\eta_{y}^{2}\sin^{2}k_{y})\bigr)^\frac{1}{2}
\end{align}
\end{widetext}

As mentioned above, the phase diagrams seen in Fig.\ \ref{pda} were mapped out by solving these equations self-consistently over the desired region of parameter space. That is to say, an initial set of mean field values were substituted into the equations and then checked against the LHS of each equation. If the difference was greater than some tolerance set at the beginning, the new values were fed back in until convergence was achieved. 
\subsection{Brillouin Zone}

The underlying bravais lattice is spanned by the primitive vectors $\left(3{\bf a}_{1},3{\bf a}_{2}\right)=\left({\bf A}_{1},{\bf A}_{2}\right)$
where ${\bf A}_{1}=\frac{3}{2}\left(\sqrt{3}\hat{x}+\hat{y}\right)a_{0}$, ${\bf A}_{2}=3\hat{y}a_{0}$ and $a_{0}$ is the lattice spacing. The reciprocal lattice vectors, ${\bf G}_{1}=\frac{2\pi}{a_{0}}\frac{2}{3\sqrt{3}}\hat{x}$ and ${\bf G}_{2}=\frac{2\pi}{a_{0}}\frac{1}{3\sqrt{3}}\left(\sqrt{3}\hat{y}-\hat{x}\right)$, then follow and we have for the brillouin zone

\begin{equation}
\begin{split}
 {\bf k}&=m{\bf G}_{1}+n{\bf G}_{2} \quad m,n\,\epsilon \ \mathbb N \\
 &=\frac{2\pi}{3\sqrt{3}a_{0}}\frac{1}{L}\left[\left(2m-n\right)\hat{x}+\sqrt{3}n\hat{y}\right]
 \end{split}
\end{equation}

Thus with  ${\bf a}_{1}=\left(-\frac{\sqrt{3}}{2}\hat{x}+\frac{1}{2}\hat{y}\right)a_{0}$,  ${\bf a}_{2}=\left(\frac{\sqrt{3}}{2}\hat{x}+\frac{1}{2}\hat{y}\right)a_{0}$ and  ${\bf a}_{3}=-\hat{y}a_{0}$ 
we get, $k_{1}={\bf a}_{1} \cdot {\bf k}=\frac{2\pi}{3}\frac{n-m}{L}$, $k_{2}={\bf a}_{2} \cdot {\bf k}=\frac{2\pi}{3}\frac{m}{L}$ 
and $k_{3}={\bf a}_{3} \cdot {\bf k}=-\frac{2\pi}{3}\frac{n}{L}$.


\begin{thebibliography}{10}

\bibitem{graphene} A. H. Castro Neto, F. Guinea, N. M. Peres, K. S. Novoselov, and A. K. Geim, \rmp {\bf 81}, 109 (2009). 

\bibitem{berciu1} M.~Berciu,T.G.~Rappoport,  and B.~Janko,
 {\em Nature} {\bf 435}, 71-75 (2005). 

\bibitem{lanzara1} S.Y. Zhou, G.-H. Gweon, A.V. Fedorov, P.N. First, W.A. de Heer, D.-H. Lee, F. Guinea, A.H. Castro Neto, A. Lanzara, Nature Mat. {\bf 6}, 770-775 (2007).

\bibitem{haldane1} F.D.M. Haldane, \prl {\bf 61}, 2015 (1988).

\bibitem{zhang1} S. Murakami, N. Nagaosa, and S.-C. Zhang, \prl {\bf 93}, 156804, (2004).

\bibitem{kane1} C.~L. Kane, and E.~J. Mele, \prl {\bf 95}, 146802 (2005).

\bibitem{tknn}  D. J. Thouless, M. Kohmoto, M. P. Nightingale, and M. den Nijs, \prl {\bf 49}, 405 (1982).

\bibitem{kane2} C.~L. Kane, and E.~J. Mele, \prl {\bf 95}, 226801 (2005).

\bibitem{moore1} J.~E. Moore and L.~Balents, \prb {\bf 75} 121306(R) (2007). 

\bibitem{bernevig1} B.A.~Bernevig, T.L.~Hughes, and S.-C.~Zhang,
  Science {\bf 314}, 1757 (2006).

\bibitem{konig1}
M. K\"{o}nig {\em et al.}, Science {\bf 318}, 766 (2008). 

\bibitem{roy1} R.~Roy, \prb {\bf 79}, 195322 (2009).

\bibitem{fu1} 
L.~Fu, C.~L. Kane, and E.~J. Mele, \prl {\bf 98} 106803 (2007).

\bibitem{HouChaMud07a}
C.-Y.~Hou, C.~Chamon, and C.~Mudry,
\newblock \prl {\bf 98}, 186809 (2007).

\bibitem{SerFra07x}
B.~Seradjeh and M.~Franz,
\newblock \prl {\bf 101}, 146401 (2008).

\bibitem{ryu1} S. Ryu, C.~Mudry, C.-Y.~Hou, and C.~Chamon,
arXiv:0908.3054

\bibitem{Ser08a}
B.~Seradjeh, C. ~Weeks, and M.~Franz,
\newblock \prb {\bf 77}, 033104 (2008).

\bibitem{Raghu} S.~Raghu, X.-L. Qi,C.~Honerkamp, and S.-C.~Zhang,
\newblock \prl {\bf 100}, 156401 (2008).

\bibitem{drut1} J.E.~Drut and T.A.~L\"ahde, \prl {\bf 102}, 026802 (2009).

\bibitem{du1}
X.~Du, I.~Skachko, F.~Duerr, A.~Luican, and  E.Y.~Andrei, Nature advance online publication 14 October 2009 | doi:10.1038/nature08522.

\bibitem{Hou2} C.-Y.~Hou, C.~Chamon, and C.~Mudry, arXiv:0909.2984


\bibitem{guo1} H.-M.~Guo and M.~Franz, \prb {\bf 80}, 113102 (2009).

\bibitem{mallet1} P.~Mallet, F.~Varchon,  C.~Naud, L.~Magaud, C.~Berger, and J.-Y.~Veuillen, \prb {\bf 76}, 041403(R) (2007).

\bibitem{tami1} T.~Pereg-Barnea and A.H.~MacDonald, \prb {\bf 78}, 014201 (2008), and references therein.

\end{thebibliography}
\end{document}